\documentstyle[10pt,epsfig,dp_delphititle]{dp_delphi}
\makeindex
\def\DpPaperGroup{PH-EP}
\def\DpPaperRef{2007-012}
\def\DpDate{25 May 2007}
\def\DpAuthors{DELPHI Collaboration}
\def\DpSubmit{(Accepted by Eur. Phys. J. C)}
\def\DpTitle{
Observation of the Muon Inner Bremsstrahlung at LEP1}
\def\DpComment{}
\def\DpEMail{}

\newcommand{\be}{\begin{equation}}
\newcommand{\ee}{\end{equation}}
\newcommand{\bd}{\begin{displaymath}}
\newcommand{\ed}{\end{displaymath}}
\newcommand{\bt}{\begin{tabular}}
\newcommand{\et}{\end{tabular}}
\newcommand{\efig}{\end{figure}}
\newcommand{\bc}{\begin{center}}
\newcommand{\ec}{\end{center}}

\begin{document}
\makeatletter
\makeatother

\begin{titlepage}
\pagenumbering{roman}

\CERNpreprint{\DpPaperGroup}{\DpPaperRef}   
\date{{\small\DpDate}}			    
\title{\DpTitle}			    
\address{\DpAuthors}			    

\begin{shortabs}			    
\noindent
%
\noindent
Muon bremsstrahlung photons converted in front of the DELPHI main tracker
(TPC) in dimuon events at LEP1 were studied in two photon kinematic ranges:
$0.2 < E_\gamma \leq 1$ GeV and transverse momentum with respect to the
parent muon $p_T < 40$ MeV/$c$, and $1 < E_\gamma \leq 10$ GeV and
$p_T < 80$ MeV/$c$ . A good agreement of the observed photon rate
with predictions from QED for the muon inner bremsstrahlung was found,
contrary to the anomalous soft photon excess that has been observed
recently in hadronic $Z^0$ decays.
The obtained ratios of the observed signal to the predicted level of the
muon bremsstrahlung are $1.06 \pm 0.12 \pm 0.07$ in the photon energy range
$0.2 < E_\gamma \leq 1$ GeV and $1.04 \pm 0.09 \pm 0.12$ in the photon energy
range $1 < E_\gamma \leq 10$ GeV. The bremsstrahlung dead cone is observed
for the first time in the direct photon production at LEP.
\end{shortabs}

\vfill

\begin{center}
\DpSubmit \ \\          
\DpComment \ \\
\DpEMail \ \\
\end{center}

\vfill
\clearpage

\headsep 10.0pt

\addtolength{\textheight}{10mm}
\addtolength{\footskip}{-5mm}

\begingroup
%
\newcommand{\DpName}[2]{\hbox{#1$^{\ref{#2}}$},\hfill}
\newcommand{\DpNameTwo}[3]{\hbox{#1$^{\ref{#2},\ref{#3}}$},\hfill}
\newcommand{\DpNameThree}[4]{\hbox{#1$^{\ref{#2},\ref{#3},\ref{#4}}$},\hfill}
\newskip\Bigfill \Bigfill = 0pt plus 1000fill
\newcommand{\DpNameLast}[2]{\hbox{#1$^{\ref{#2}}$}\hspace{\Bigfill}}
%
\footnotesize
\noindent
\DpName{J.Abdallah}{LPNHE}
\DpName{P.Abreu}{LIP}
\DpName{W.Adam}{VIENNA}
\DpName{P.Adzic}{DEMOKRITOS}
\DpName{T.Albrecht}{KARLSRUHE}
\DpName{R.Alemany-Fernandez}{CERN}
\DpName{T.Allmendinger}{KARLSRUHE}
\DpName{P.P.Allport}{LIVERPOOL}
\DpName{U.Amaldi}{MILANO2}
\DpName{N.Amapane}{TORINO}
% ...
\normalsize
\endgroup
\newpage
\titlefoot{Department of Physics and Astronomy, Iowa State
     University, Ames IA 50011-3160, USA
    \label{AMES}}
\titlefoot{IIHE, ULB-VUB,
     Pleinlaan 2, B-1050 Brussels, Belgium
    \label{BRUSSELS}}
\titlefoot{Physics Laboratory, University of Athens, Solonos Str.
     104, GR-10680 Athens, Greece
    \label{ATHENS}}
\titlefoot{Department of Physics, University of Bergen,
     All\'egaten 55, NO-5007 Bergen, Norway
    \label{BERGEN}}
\titlefoot{Dipartimento di Fisica, Universit\`a di Bologna and INFN,
     Via Irnerio 46, IT-40126 Bologna, Italy
    \label{BOLOGNA}}
\titlefoot{Centro Brasileiro de Pesquisas F\'{\i}sicas, rua Xavier Sigaud 150,
     BR-22290 Rio de Janeiro, Brazil
    \label{BRASIL-CBPF}}
\titlefoot{Inst. de F\'{\i}sica, Univ. Estadual do Rio de Janeiro,
     rua S\~{a}o Francisco Xavier 524, Rio de Janeiro, Brazil
    \label{BRASIL-IFUERJ}}
\titlefoot{Coll\`ege de France, Lab. de Physique Corpusculaire, IN2P3-CNRS,
     FR-75231 Paris Cedex 05, France
    \label{CDF}}
\titlefoot{CERN, CH-1211 Geneva 23, Switzerland
    \label{CERN}}
\titlefoot{Institut de Recherches Subatomiques, IN2P3 - CNRS/ULP - BP20,
     FR-67037 Strasbourg Cedex, France
    \label{CRN}}
\titlefoot{Now at DESY-Zeuthen, Platanenallee 6, D-15735 Zeuthen, Germany
    \label{DESY}}
\titlefoot{Institute of Nuclear Physics, N.C.S.R. Demokritos,
     P.O. Box 60228, GR-15310 Athens, Greece
    \label{DEMOKRITOS}}
\titlefoot{FZU, Inst. of Phys. of the C.A.S. High Energy Physics Division,
     Na Slovance 2, CZ-182 21, Praha 8, Czech Republic
    \label{FZU}}
\titlefoot{Dipartimento di Fisica, Universit\`a di Genova and INFN,
     Via Dodecaneso 33, IT-16146 Genova, Italy
    \label{GENOVA}}
\titlefoot{Institut des Sciences Nucl\'eaires, IN2P3-CNRS, Universit\'e
     de Grenoble 1, FR-38026 Grenoble Cedex, France
    \label{GRENOBLE}}
\titlefoot{Helsinki Institute of Physics and Department of Physical Sciences,
     P.O. Box 64, FIN-00014 University of Helsinki, 
     \indent~~Finland
    \label{HELSINKI}}
\titlefoot{Joint Institute for Nuclear Research, Dubna, Head Post
     Office, P.O. Box 79, RU-101 000 Moscow, Russian Federation
    \label{JINR}}
\titlefoot{Institut f\"ur Experimentelle Kernphysik,
     Universit\"at Karlsruhe, Postfach 6980, DE-76128 Karlsruhe,
     Germany
    \label{KARLSRUHE}}
\titlefoot{Institute of Nuclear Physics PAN,Ul. Radzikowskiego 152,
     PL-31142 Krakow, Poland
    \label{KRAKOW1}}
\titlefoot{Faculty of Physics and Nuclear Techniques, University of Mining
     and Metallurgy, PL-30055 Krakow, Poland
    \label{KRAKOW2}}
\titlefoot{Universit\'e de Paris-Sud, Lab. de l'Acc\'el\'erateur
     Lin\'eaire, IN2P3-CNRS, B\^{a}t. 200, FR-91405 Orsay Cedex, France
    \label{LAL}}
\titlefoot{School of Physics and Chemistry, University of Lancaster,
     Lancaster LA1 4YB, UK
    \label{LANCASTER}}
\titlefoot{LIP, IST, FCUL - Av. Elias Garcia, 14-$1^{o}$,
     PT-1000 Lisboa Codex, Portugal
    \label{LIP}}
\titlefoot{Department of Physics, University of Liverpool, P.O.
     Box 147, Liverpool L69 3BX, UK
    \label{LIVERPOOL}}
\titlefoot{Dept. of Physics and Astronomy, Kelvin Building,
     University of Glasgow, Glasgow G12 8QQ, UK
    \label{GLASGOW}}
\titlefoot{LPNHE, IN2P3-CNRS, Univ.~Paris VI et VII, Tour 33 (RdC),
     4 place Jussieu, FR-75252 Paris Cedex 05, France
    \label{LPNHE}}
\titlefoot{Department of Physics, University of Lund,
     S\"olvegatan 14, SE-223 63 Lund, Sweden
    \label{LUND}}
\titlefoot{Universit\'e Claude Bernard de Lyon, IPNL, IN2P3-CNRS,
     FR-69622 Villeurbanne Cedex, France
    \label{LYON}}
\titlefoot{Dipartimento di Fisica, Universit\`a di Milano and INFN-MILANO,
     Via Celoria 16, IT-20133 Milan, Italy
    \label{MILANO}}
\titlefoot{Dipartimento di Fisica, Univ. di Milano-Bicocca and
     INFN-MILANO, Piazza della Scienza 3, IT-20126 Milan, Italy
    \label{MILANO2}}
\titlefoot{IPNP of MFF, Charles Univ., Areal MFF,
     V Holesovickach 2, CZ-180 00, Praha 8, Czech Republic
    \label{NC}}
\titlefoot{NIKHEF, Postbus 41882, NL-1009 DB
     Amsterdam, The Netherlands
    \label{NIKHEF}}
\titlefoot{National Technical University, Physics Department,
     Zografou Campus, GR-15773 Athens, Greece
    \label{NTU-ATHENS}}
\titlefoot{Physics Department, University of Oslo, Blindern,
     NO-0316 Oslo, Norway
    \label{OSLO}}
\titlefoot{Dpto. Fisica, Univ. Oviedo, Avda. Calvo Sotelo
     s/n, ES-33007 Oviedo, Spain
    \label{OVIEDO}}
\titlefoot{Department of Physics, University of Oxford,
     Keble Road, Oxford OX1 3RH, UK
    \label{OXFORD}}
\titlefoot{Dipartimento di Fisica, Universit\`a di Padova and
     INFN, Via Marzolo 8, IT-35131 Padua, Italy
    \label{PADOVA}}
\titlefoot{Rutherford Appleton Laboratory, Chilton, Didcot
     OX11 OQX, UK
    \label{RAL}}
\titlefoot{Dipartimento di Fisica, Universit\`a di Roma II and
     INFN, Tor Vergata, IT-00173 Rome, Italy
    \label{ROMA2}}
\titlefoot{Dipartimento di Fisica, Universit\`a di Roma III and
     INFN, Via della Vasca Navale 84, IT-00146 Rome, Italy
    \label{ROMA3}}
\titlefoot{DAPNIA/Service de Physique des Particules,
     CEA-Saclay, FR-91191 Gif-sur-Yvette Cedex, France
    \label{SACLAY}}
\titlefoot{Instituto de Fisica de Cantabria (CSIC-UC), Avda.
     los Castros s/n, ES-39006 Santander, Spain
    \label{SANTANDER}}
\titlefoot{Inst. for High Energy Physics, Serpukov
     P.O. Box 35, Protvino, (Moscow Region), Russian Federation
    \label{SERPUKHOV}}
\titlefoot{J. Stefan Institute, Jamova 39, SI-1000 Ljubljana, Slovenia
    \label{SLOVENIJA1}}
\titlefoot{Laboratory for Astroparticle Physics,
     University of Nova Gorica, Kostanjeviska 16a, SI-5000 Nova Gorica, Slovenia
    \label{SLOVENIJA2}}
\titlefoot{Department of Physics, University of Ljubljana,
     SI-1000 Ljubljana, Slovenia
    \label{SLOVENIJA3}}
\titlefoot{Fysikum, Stockholm University,
     Box 6730, SE-113 85 Stockholm, Sweden
    \label{STOCKHOLM}}
\titlefoot{Dipartimento di Fisica Sperimentale, Universit\`a di
     Torino and INFN, Via P. Giuria 1, IT-10125 Turin, Italy
    \label{TORINO}}
\titlefoot{INFN,Sezione di Torino and Dipartimento di Fisica Teorica,
     Universit\`a di Torino, Via Giuria 1,
     IT-10125 Turin, Italy
    \label{TORINOTH}}
\titlefoot{Dipartimento di Fisica, Universit\`a di Trieste and
     INFN, Via A. Valerio 2, IT-34127 Trieste, Italy
    \label{TRIESTE}}
\titlefoot{Istituto di Fisica, Universit\`a di Udine and INFN,
     IT-33100 Udine, Italy
    \label{UDINE}}
\titlefoot{Univ. Federal do Rio de Janeiro, C.P. 68528
     Cidade Univ., Ilha do Fund\~ao
     BR-21945-970 Rio de Janeiro, Brazil
    \label{UFRJ}}
\titlefoot{Department of Radiation Sciences, University of
     Uppsala, P.O. Box 535, SE-751 21 Uppsala, Sweden
    \label{UPPSALA}}
\titlefoot{IFIC, Valencia-CSIC, and D.F.A.M.N., U. de Valencia,
     Avda. Dr. Moliner 50, ES-46100 Burjassot (Valencia), Spain
    \label{VALENCIA}}
\titlefoot{On leave of absence from ITEP, Moscow, Russian Federation
    \label{ITEP}}
\titlefoot{Institut f\"ur Hochenergiephysik, \"Osterr. Akad.
     d. Wissensch., Nikolsdorfergasse 18, AT-1050 Vienna, Austria
    \label{VIENNA}}
\titlefoot{Inst. Nuclear Studies and University of Warsaw, Ul.
     Hoza 69, PL-00681 Warsaw, Poland
    \label{WARSZAWA}}
\titlefoot{Now at University of Warwick, Coventry CV4 7AL, UK
    \label{WARWICK}}
\titlefoot{Fachbereich Physik, University of Wuppertal, Postfach
     100 127, DE-42097 Wuppertal, Germany \\
\noindent
{$^\dagger$~deceased}
    \label{WUPPERTAL}}
\addtolength{\textheight}{-10mm}
\addtolength{\footskip}{5mm}
\clearpage

\headsep 30.0pt
\end{titlepage}

%
\pagenumbering{arabic}				    
\setcounter{footnote}{0}			    %
\large
%
\section{Introduction}
Recent observation of anomalous soft photon production in hadronic  
$Z^0$ decays collected in the DELPHI experiment at LEP1 \cite{aspdel}
has demonstrated the persistence of the soft photon anomaly found earlier
in several fixed target experiments with high energy hadronic beams, 
\cite{wa27,na22,wa83,wa91,wa102}. 
The photon kinematic range was defined in \cite{aspdel} as follows: 
$0.2 < E_{\gamma} \leq 1$ GeV, $p_T < 80$ MeV/$c$, the $p_T$ being the photon 
transverse momentum with respect to the parent jet direction. Though the 
reaction $e^+e^- \rightarrow Z^0 \rightarrow hadrons$ presents a distinct 
mechanism of hadron production as compared to \cite{wa27,na22,wa83,wa91,wa102},
the observed soft photon production characteristics were found in 
\cite{aspdel} to be very close to those reported in 
\cite{wa27,na22,wa83,wa91,wa102}, 
both for the measured production rate and for the observed
ratio of the rate to the inner hadronic bremsstrahlung. The latter was 
expected to be the main source of the direct soft photons in kinematic 
ranges under study (see \cite{lan,low,gribov}), while 
the observed signals were found in \cite{aspdel,wa27,na22,wa83,wa91,wa102}
to be several times higher than the bremsstrahlung predictions. No 
theoretical explanation of this excess is available so far; reviews of 
the theoretical approaches to the problem can be found in \cite{pis,lich} 
(see also the references [13-33] in \cite{aspdel}). 
   
From the experimental analysis, given a similarity of the soft photon 
production characteristics in both classes of experiments, the conclusion 
was drawn in \cite{aspdep} that the excess photons are created 
during the process of hadronization of quarks, 
i.e. their origin is strongly restricted to reactions of hadron production. 
If this ansatz is correct, a good agreement should be found between theory 
and experiment for the direct soft photon production in reactions of pure 
electroweak nature. What is the experimental situation in this field?

The electron inner bremsstrahlung in $e^+ e^-$ collisions (initial state 
radiation, ISR) was an important (and rather inconvenient) effect at LEP, with
which all the LEP experiments had to contend. No deviation of the ISR 
characteristics from those expected from theory was observed, either at $Z^0$
or at high energy (see for example the DELPHI studies \cite{isr}). Therefore 
the situation with the electron inner bremsstrahlung can be considered as 
showing a nice agreement between theory and experiment. 

On the other hand, tests of QED with the muon inner bremsstrahlung
which appears as final state radiation (FSR) in 
$e^+ e^- \rightarrow \mu^+ \mu^-$ events were scarce at LEP. There were only 
two studies of photon production in $Z^0 \rightarrow \mu^+ \mu^-$ events at 
LEP1 \cite{delphimu,opalmu} and a single study of 
$e^+ e^- \rightarrow \mu^+ \mu^-$ events at LEP2 \cite{delphimu2} 
\footnote{Outside the LEP experiments, 
a few studies of the muon inner bremsstrahlung have been done, see 
\cite{mubneu} and references therein.}. All these studies aimed at the 
separation of rather hard photons, isolated from the neighbouring muon. 
So, the DELPHI analysis of final state radiation from muons at LEP1
\cite{delphimu} was restricted to the photon kinematic range of
$E_{\gamma} > 2$ GeV, $\theta_{\mu\gamma} > 5^{\circ}$, i.e. to the
transverse momenta with respect to the muon direction $p_T > 174$ MeV/$c$.
In \cite{delphimu2} the minimum value of the angle $\theta_{\mu\gamma}$ 
was increased up to $15^{\circ}$ (keeping the same photon energy 
threshold), tripling the minimum photon $p_T$. 
The OPAL analysis at LEP1 \cite{opalmu} used photons of $E_{\gamma} > 0.9$ 
GeV and $\theta_{\mu\gamma} > 200$ mrad, i.e. the photon transverse 
momenta with respect to the muon direction were $p_T > 179$ MeV/$c$.
Thus, an analysis of the muon inner bremsstrahlung in the soft photon
kinematic range close to that analyzed in \cite{aspdel} is completely
missing at LEP. This motivated us to study the reaction
\begin{equation}
e^+e^- \rightarrow Z^0 \rightarrow \mu^+ \mu^- n\gamma, ~~~~~~~~~n\geq 1
\end{equation}
at LEP1 in a photon kinematic range similar to the one analyzed in 
\cite{aspdel} (with the photon transverse momentum being defined now with 
respect to the parent muon direction). In addition to the low 
energy (LE) band of $0.2 < E_{\gamma} \leq 1$ GeV explored in \cite{aspdel}, 
a higher energy (HE) band  of $ 1 < E_{\gamma} \leq 10$ GeV was also used 
in the analysis, being restricted however to the photons of small transverse
momentum with respect to the parent muon direction, $p_T < 80$ MeV/$c$. 
The $p_T$ range of the LE band chosen for the definition of the bremsstrahlung 
signal was taken narrower in this work as compared to that in \cite{aspdel}, 
namely $p_T < 40$ MeV/$c$. This choice was motivated by the fact that the
photon angular variable used in this analysis, the photon polar angle 
relative to the parent muon direction, can be measured much more accurately 
as compared to the angular variable used in \cite{aspdel}, the photon
polar angle relative to the parent jet direction, and this confined most of 
the LE bremsstrahlung photons down to the mentioned $p_T$ range.  

The results obtained in this study are presented both uncorrected and
corrected for the photon detection efficiency. The presentation of the
uncorrected results is motivated by their better statistical accuracies
and smaller systematic uncertainties in the absolute photon rates.

\section{ Theoretical predictions for the muon inner bremsstrahlung}   
In electroweak reactions like (1) the inner bremsstrahlung is a process of 
direct photon production calculated via purely QED machinery.
The production rates for the bremsstrahlung photons from 
colliding $e^+ e^-$ (ISR) and from final $\mu^+ \mu^-$ pairs (FSR) 
in the $p_T$ range under study can be calculated at once using a universal 
formula descending from Low \cite{low} with a modification suggested
by Haissinski \cite{hais}:

\begin{equation}
\frac{dN_{\gamma}}{d^{3}\vec{k}}
=
\frac{\alpha}{(2 \pi)^2} \frac{1}{E_\gamma}
\int d^3 \vec{p}_{\mu}  
\sum_{i,j} \eta_{i} \eta_{j}
\frac{(\vec{p}_{i \bot} \cdot \vec{p}_{j \bot}) }{ ( P_{i} K )  ( P_{j} K )}
\frac{ d N_{\mu}}{ d^{3} \vec{p}_{\mu} } 
\end{equation}
                                                                                
\noindent
where $K$ and $\vec{k}$ denote photon four- and three-momenta, 
$P$ are the 4-momenta of the beam $e^+, e^-$ and the muon involved, 
and $\vec{p}_\mu$ is the 3-momentum of the muon;
$\vec{p}_{i \bot} = \vec{p}_i-(\vec{n} \cdot \vec{p}_i) \cdot \vec{n}$ and
~$\vec{n}$ is the photon unit vector, $\vec{n} = \vec{k}/k $;
$\eta=1$ for the beam electron and for the outgoing $\mu^+$,
$\eta=-1$ for the beam positron and for the outgoing $\mu^-$,
and the sum is extended over both beam particles and the parent muon (formula 
(2) is presented in the form of the photon production rate per muon);
the last factor in the integrand is a differential production rate of the
parent muon.

As can be seen, formula (2) is of the lowest (leading) order in $\alpha$.
Higher order radiative corrections to it can be evaluated using exponentiated
photon spectra in the LE and HE bands. In the accepted regions of low $p_T$
the effects of the exponentiation were found to be rather small, 
as considered in Sect. 6.2. 

To a great extent, formula (2) is used in this paper specifically to enable a
comparison with the corresponding formula applied for the calculation of
the inner hadronic bremsstrahlung in hadronic decays of $Z^0$ \cite{aspdel} 
(cf. the analogous formulae in \cite{wa27,na22,wa83,wa91,wa102}):
                                                                             
\begin{equation}
\frac{dN_{\gamma}}{d^{3}\vec{k}}
=
\frac{\alpha}{(2 \pi)^2} \frac{1}{E_\gamma}
\int d^3 \vec{p}_{1} . . . d^3 \vec{p}_{N}
\sum_{i,j} \eta_{i} \eta_{j}
\frac{(\vec{p}_{i \bot} \cdot \vec{p}_{j \bot}) }{ ( P_{i} K )  ( P_{j} K )}
\frac{ d N_{h}}{ d^{3} \vec{p}_{1} ... d^{3} \vec{p}_{N}}
\end{equation}
                                                                                
\noindent
where $K$ and $\vec{k}$ denote again photon four- and three-momenta, 
$P$ are the 4-momenta of the beam $e^+, e^-$ and $N$ charged outgoing 
hadrons, and $\vec{p}_1$ ... $\vec{p}_N$ are the 3-momenta of the hadrons;
$\eta=1$ for the beam electron and for positive outgoing hadrons,
$\eta=-1$ for the beam positron and negative outgoing hadrons,
and the sum is extended over all the $N+2$ charged particles involved;
the last factor in the integrand is a differential hadron production rate
(when calculating the photon production rate per jet only hadrons 
lying in the forward hemisphere of  
a given jet enter the sum). Calculations performed with formulae (2,3) 
show that the inner bremsstrahlung rate from one muon is approximately equal, 
in the kinematic region under study, to the predicted inner hadronic 
bremsstrahlung from a whole hadronic jet of a $Z^0$ hadronic decay. 
To a great extent, this is a consequence of the coherence
of the photon radiation from the individual radiation sources, the charged 
hadrons produced in the fragmentation process.
 
The contribution of the ISR to these rates is small, being below 1\% 
in the photon kinematic range chosen for the analysis.
This smallness is easy to understand: although the
ISR from electron/positron beams is much more intense than the ISR from
hadron beams in experiments \cite{wa27,na22,wa83,wa91,wa102}, where it 
contributed a significant amount to the detected photon rate, all 
the extra photons in an experiment with colliding $e^+ e^-$ are emitted at 
very small polar angles with respect to the beam directions, with the angular 
distribution peaking at $\Theta_{\gamma} = \sqrt{3}/\Gamma$, where $\Gamma$ 
is a beam Lorentz factor ($\Gamma = 0.89 \times 10^5$ at the $Z^0$ peak), 
thus yielding few photons in the barrel region used in our analysis.

The muon bremsstrahlung radiation (FSR) has the same angular behaviour
of the photon production rate versus the photon polar angle 
relative to the parent muon direction (the photon production angle, 
$\theta_\gamma$), with $\Gamma$ being in this case a muon Lorentz factor. 
For the muons from $Z^0$ decays at rest the $\Gamma = 4.3 \times 10^2$ 
corresponds to the peak position at 4.0 mrad. 
Note that the position of the peak does not depend on the bremsstrahlung photon 
energy, since the dependences of the photon production rate on the photon 
energy and the photon production angle are factorized in formulae (2,3). 
The turnover of the muon bremsstrahlung angular distribution at the peak 
value and its vanishing at $\theta_{\gamma} \rightarrow 0$  
is termed the dead cone effect. This behaviour is illustrated by Fig. 1a
where the initial part of the production angle distribution for the FSR
of the reaction (1) is shown, generated \footnote{The Monte Carlo data set 
of dimuon events described below was used as the input of the generation.}
with formula (2). The observation of the dead cone presents an experimental 
challenge requiring a highly accurate apparatus; the angular resolution of 
the opening angle between the measured muon and photon directions which is 
necessary for the observation of the muon bremsstrahlung dead cone  
at LEP1, has to be of the order of 1$-$2 mrad.

\section{Experimental technique}   
\subsection{The DELPHI detector}
The DELPHI detector is described in detail elsewhere \cite{delphi1,delphi2}.
The following is a brief description of the subdetector units relevant
for this analysis. 

In the DELPHI reference frame the $z$ axis is taken
along the direction of the $e^-$ beam. The angle $\Theta$ is the polar
angle defined with respect to the $z$-axis, $\Phi$ is the azimuthal angle
about this axis and $R$ is the distance from this axis.
                                                                                
The TPC, the principal device used in this analysis, was the main
tracker of the DELPHI detector; it covered the angular range from
$20^{\circ}$ to $160^{\circ}$ in $\Theta$ and extended
from 30 cm to 122 cm in R. It provided up to 16 space points for
pattern recognition and ionization information extracted from 192 wires.
The TPC together with other tracking devices (Vertex Detector, Inner Detector, 
Outer Detector and Forward Chambers) ensured a very good angular accuracy of 
the muon track reconstruction, which is a part of the overall angular 
resolution for the photon production angle. The distribution of the opening 
angles between the generated and reconstructed muon directions is
shown in Fig. 1b; it can be characterized by the distribution mean of 
0.42 mrad and its r.m.s. width of 0.37 mrad, which restricts 90\% of
the entries within the 0$-$1 mrad interval. 
                                                                                
The identification of muons was based on the muon chambers (MUC) surrounding 
the detector, the hadron calorimeter (HCAL) and the electromagnetic 
calorimeter (High density Projection Chamber, HPC), as described in 
\cite{muid}.

The Monte Carlo (MC) data set used in this analysis was produced with the
DYMU3 generator \cite{dymu3}. Higher order radiative corrections to the 
reaction (1) total cross section were accounted for via the exponentiation 
procedure implemented in the generator. The generated dimuon events were 
passed through the DELPHI detector simulation program DELSIM \cite{delphi2}. 

\subsection{Detection of photons}
Photon conversions in front of the main DELPHI tracker (TPC) were reconstructed
by an algorithm that examined the tracks reconstructed in the TPC. A search 
was made along each TPC track for the point where the tangent of the track 
trajectory points directly to the beam spot in the $R\Phi$ projection.
Under the assumption that the opening angle of the electron-positron pair
is zero, this point represented a possible photon conversion point at
radius $R$. All tracks which have had a solution $R$ that was more than one
standard deviation away from the main vertex, as defined by the beam spot,
were considered to be conversion candidates. If two oppositely charged
conversion candidates were found with compatible conversion point
parameters they were linked together to form the converted photon. The
following selection criteria were imposed:
\begin{itemize}
 \item the $\Phi$ difference between the two conversion points should be 
at most 30 mrad;
 \item a possible difference between the polar angles $\Theta$ of the two 
tracks should be at most 15 mrad;
 \item at least one of the tracks should have no associated hits in front
of the reconstructed mean conversion radius.
\end{itemize}
For the pairs fulfilling these criteria a $\chi^2$ was calculated from
$\Delta \Theta, \Delta \Phi$ and the difference of the reconstructed
conversion radii $\Delta R$ in order
to find the best combinations in cases where there were ambiguous
associations. A constrained fit was then applied to the electron-positron
pair candidate which forced a common conversion point with zero opening
angle and collinearity between the momentum sum and the line from the
beam spot to the conversion point.
                                                                                
The photon detection efficiency, i.e. the conversion probability combined
with the reconstruction efficiency, was determined with the hadronic MC data 
since the converted photon sample in dimuon events was insufficient 
statistically for such a determination. The efficiencies were
tabulated against three variables: $E_{\gamma}$,  $\Theta_{\gamma}$ (the
photon polar angle to the beam), and $\theta_{\gamma tk}$ (the photon opening
angle to the closest track). The efficiency varied with the energy from zero 
at the 0.2 GeV detection threshold up to 4 - 6\% at $E_{\gamma} \geq 1$ GeV, 
depending on the two other variables (for details see \cite{aspdel}). 

In order to reduce a possible difference in the reconstruction of the 
converted photons in the MC and the real data (originating from the bias in 
the detector material distributions in the two data sets and from a possible 
distinction in their pattern recognition results) the recalibration procedure 
described in \cite{aspdel} was implemented, with the recalibration 
coefficients obtained with hadronic events.

The angular precision of the photon direction reconstruction was studied  
using the dimuon MC events and was found to be of a Breit-Wigner shape, as 
expected for the superposition of many Gaussian distributions of varying width 
\cite{eadie}. The full widths ($\Gamma$'s) of the $\Delta \Theta_{\gamma}$ and
$\Delta \Phi_{\gamma}$ distributions were $2.3 \pm 0.1$ mrad and $1.9 \pm 0.1$
mrad, respectively, for the combined 0.2$-$10 GeV interval (Figs. 1c, 1d). The 
full width of the distribution of the difference  $\Delta \theta_{\gamma}$
between the generated and reconstructed muon-photon opening angles (which is
the difference in the production angle $\theta_{\gamma}$ defined in Sect. 2 and 
therefore represents the overall angular resolution of the current analysis) 
was found to be $2.1 \pm 0.1$ mrad (Fig. 1e), thus providing a possibility 
for the observation of the muon bremsstrahlung dead cone. Moreover,
one can improve essentially this raw resolution, 
though at the price of a loss of 50\% of the converted photon statistics, by
requiring the photon energy to exceed 1 GeV and the conversion radius to be 
greater than 25 cm. With these tighter cuts 1.4 mrad resolution (the full 
width) was achieved and used in a particular case which required a high angular 
resolution and is described below (Sect. 7.3). 

The accuracy of the converted photon energy measurement was studied also
with the dimuon MC events. In both energy bands it was at the level of  
$\pm$1.5\% (the Breit-Wigner full width about 3\%); this is illustrated by 
Fig.1f where the distribution of the relative difference between the generated 
and reconstructed photon energy is plotted for the LE photons. The resolution
was checked with events of the (hadronic) real data by comparing 
the $\pi^0$ peak width of the $\gamma \gamma$ mass distribution from these data
to the analogous one from the MC.  

\section{Data selection}   
\subsection{ Selection of dimuon events}   
The data selection was done under standard cuts aimed at the separation 
of dimuon events (cf. \cite{delphimu,muid}) which are described below. 
The consecutive application of these cuts reduced the MC sample of dimuon 
events by factors indicated in parentheses:

\begin{itemize}
 \item the number of charged particles $N_{ch}$ had to be within the interval
of $2 \leq N_{ch} \leq 5$, and the two highest momentum particles  
had to have $p > 15$ GeV/$c$ ~~(0.894);
 \item the polar angles of the two highest momentum particles had to be 
within the interval of 20$^{\circ} \leq \Theta \leq 160^{\circ}$ ~~(0.962);  
 \item the impact parameters of the two highest momentum particles had to be 
less than 0.2 and 4.5 cm in the $R\Phi$ and $z$ projections, 
respectively ~~(0.993);
 \item no additional charged particles with momenta greater than 10 GeV/$c$ 
were allowed, unless the fastest particle had a momentum greater than 
40 GeV/$c$ ~~(0.999);
 \item  the acollinearity of the two highest momentum particles had to be
less than $10^{\circ}$ ~~(0.989);
 \item  the two highest momentum particles had to be identified as muons using 
either the muon chambers (MUC), the hadron calorimeter (HCAL), or the
electromagnetic calorimeter (HPC), by requiring associated hits in the muon 
chambers, or by energy deposition in the calorimeters consistent with 
a minimum ionizing particle  ~~(0.825).

\end{itemize}

The total reduction factor for the MC events was 0.696. 

A total of 122 812 events of real data (RD) was selected under these cuts 
and compared to 373 918 selected MC events
corresponding, after the normalization of the equivalent MC luminosity 
to the integrated RD luminosity, to 121 000 expected events.


\subsection{ Selection of photons}   
The standard selection of converted photons was done under the following cuts:
\begin{itemize}
 \item only converted photons with both $e^+, e^-$ arms reconstructed were
considered;
 \item 20$^{\circ} \leq \Theta_{\gamma} \leq 160^{\circ}$;
 \item 5 cm $\leq R_{conv} \leq 50$ cm, where $R_{conv}$ means conversion 
radius;
 \item 200 MeV $ < E_{\gamma} \leq 10$ GeV.
\end{itemize}

384 and 1097 converted photons were found using these cuts in the real data
in the LE and HE energy bands, respectively.
Of these, 127 and 265 photons are in the selected $p_T$ regions: 
$p_T <$ 40 MeV/$c$ for the LE band and $p_T <$ 80 MeV/$c$ for the HE band.

For a particular analysis done to scrutinize the dead cone effect (described
in Sect. 7.3), the photon energy was required to be between 1 and 10 GeV,
and the conversion radius to be between 25 and 50 cm.

\section{ Backgrounds}   
The following background sources within the $\mu^+ \mu^-$ event sample were 
considered:
\begin{itemize}
\item
External muon bremsstrahlung:\\
the bremsstrahlung radiation from muons when they pass through the 
material of the experimental setup.
\item
Secondary photons:\\
when a high energy photon (of any origin) generates an $e^+ e^-$ pair in the 
detector material in front of the TPC the pair particles may radiate 
(external) bremsstrahlung photons, which can enter our kinematic region.
\item
``Degraded" photons:\\
higher energy converted primary photons with degraded energy measurement
due to the secondary emission of (external) bremsstrahlung by at least one
of their electrons. 
\end{itemize}

DELSIM was invoked to reproduce these processes in the MC stream.

Collection of background photons (all dubbed as External Brems) in the
MC stream was done if any one of the following conditions was satisfied:
\begin{itemize}
\item
a given photon was absent at the event generator level, i.e. in the 
DYMU3 event record;  
\item
a given photon, found in the DYMU3 event record, migrated from outside a
selected $p_T$ region into that region due to the energy degradation. 
\end{itemize}

$26.0 \pm 2.9$ and $61.5 \pm 4.5$ background photons (normalized to the RD 
statistics) were found in the selected $p_T$ regions: $p_T <$ 40 
MeV/$c$ for the LE band and $p_T <$ 80 MeV/$c$ for the HE band, respectively.  

The background from $Z^0 \rightarrow \tau^+ \tau^-$ events was estimated 
using the MC data produced with the KORALZ 4.0 generator \cite{koralz}
and passed through the full detector simulation and the analysis procedure.
The $\tau^+ \tau^-$ contamination of the dimuon event sample was 
found to be $1536 \pm 20$ events (1.3 \% of the dimuon sample), which 
contain zero photons in the LE band and $1.3 \pm 0.6$ photons 
in the HE band, of which 0.3 photons would be in the $p_T < 80$ MeV/$c$ 
region. In what follows, this background was neglected. 

The cosmic ray background was estimated from the real data, studying
events which originated close to the interaction point, but outside the limits 
allowed for selected events. In both energy bands its contribution to the
photon rates was below 0.1\%. 

The background from $Z^0 \rightarrow e^+ e^-$ events tested with the BABAMC 
generator \cite{babamc} with the full detector simulation was found to be 
vanishingly small. The same is valid for the 4-fermion backgrounds 
$Z^0 \rightarrow e^+ e^- \mu^+ \mu^-$ and
$Z^0 \rightarrow \mu^+ \mu^- q\overline{q}$ tested with events produced with 
generators \cite{excalibur,jetset}.

\section{Systematic errors}
\subsection{Systematic uncertainties in the determination of the signal}

Since the converted photon sample in the dimuon event statistics collected 
by the DELPHI experiment during the LEP1 period was insufficient for the
determination of the photon detection efficiencies and the recalibration
coefficients, they were taken as being defined with hadronic events. 
Therefore it is worth to start the consideration of the systematic errors and 
their estimations with the uncertainties induced by these components of the
analysis as they were determined in \cite{aspdel}. 

The uncertainty due to a difference in the photon propagation and conversion 
in the detector material in the RD and its simulation in the MC, 
and analogous difference in the pattern recognition, left after the 
recalibration procedure was applied (termed in \cite{aspdel} hardware 
systematics), was studied in \cite {aspdel} and evaluated to be 0.9\% of the 
photon rate in the LE band and 2\% in the HE band. 
 
The systematic error for the photon detection efficiency\footnote{
Note that when dealing with the data uncorrected for the detection efficiency 
the efficiency error is relevant to the bremsstrahlung predictions only
(since bremsstrahlung spectra have to be convoluted with the efficiencies
in this case). On the contrary, when dealing with the corrected data the 
efficiency uncertainty has to be applied to the measured photon rates only.},
after the recalibration procedure mentioned above being applied, is a purely
instrumental effect originating from the choice of the binning of the variables
used for the efficiency parametrization, resolution effects, etc. 
In \cite{aspdel} it was found to range from 6\% to 9\% of the photon rate. 
These estimations were tested in the current study 
with the MC dimuon events by comparing the $p_T$ spectra of the DYMU3 inner 
bremsstrahlung photons transported through the DELPHI detector by DELSIM
with a subsequent simulation of their conversions, with the spectra of the 
same photons taken at the generator level and convoluted with the photon 
detection efficiency. In both energy bands the difference was below 5\% 
which was the level of the test statistical accuracy. This means that the 
aforementioned error due to the detection efficiency is likely to be 
overestimated in \cite{aspdel}, or it is really smaller in the muonic data,
in particular, due to a better angular resolution and due to a narrower angular
ranges in both energy bands. In what follows, the value of 5\% is used as an 
estimate for the uncertainty of the efficiency calculation. 

The systematic errors originating from the influence 
of the $p_T$ resolution on the selection cuts
were estimated from runs with the reconstructed photon energy and 
production angle randomly shifted according to the appropriate resolution 
function (taking into account the different angular resolutions in the LE and 
HE bands). The changes were found to be less than 0.3\% of the photon rate in 
the LE band and less than 0.4\% of the photon rate in the HE band. In what 
follows, the corresponding errors were neglected. 

The uncertainty of the background (BG) estimation is composed of the
uncertainties coming from the DYMU3 generator, efficiency and hardware 
systematics, BG selection,
and the procedure of the BG photon conversion simulation. 
The systematic error from the photon conversion simulation is considered 
to be equal to the systematic error of the photon detection in 
the MC stream, before the recalibration is applied, i.e. it can be 
approximated by the recalibration corrections, which were within 3-4\%. The 
systematic errors due to efficiency and hardware in the background estimation 
have strong positive correlations with the analogous components of the 
systematic error in the calculation of the real data photon rates (indeed
they are of the same relative amplitude, but the background errors have
to be reduced by factors of 3.9 and 3.3 in the LE and HE bands, respectively,
when entering the final systematic error, since the background rates within 
the corresponding $p_T$ intervals constitute 25.7\% and 30.2\% of the RD$-$BG 
photon rates in the corresponding energy bands). Ignoring, 
for the sake of clarity, these correlations we will 
consider all the background systematics components as independent and 
uncorrelated  with the analogous components in the RD rates. Then, calculating 
the background systematic errors similarly to those for the RD and taking into 
account the suppression factors mentioned above, the 
systematic background uncertainties appear to be 1.4\% and 2.2\% 
relative to the signal rate in the respective energy band in the case of 
the uncorrected data, and 1.9\% and 2.7\% in the case of the data corrected 
for efficiency.    

The above systematic errors are summarized in Table 1.

\vskip 0.2cm
{\bf Table 1.} Systematic uncertainties (in \% of the photon rates
in the $p_T$ ranges below 40 MeV/$c$ and 80 MeV/$c$ for the LE and HE
photons, respectively) for the signal and the predicted muon inner 
bremsstrahlung. The total systematic error of each of the two energy band 
photon rates and signal-to-bremsstrahlung ratios is the quadratic sum of 
the corresponding individual errors, as quoted in Tables 2,3 below. 
\begin{center}
\begin{tabular} { |c c c| }
\hline
&&\\
Component         & Data uncorrected for    &  Data corrected for\\
                  & the detection efficiency&  the detection efficiency\\
&&\\
\hline
&&\\
                  &  LE band~~~~~~HE band     &~~~~~LE band~~~~~~HE band~~~~~\\
&&\\
                  &~~~~~~~~~~~~~{\bf Signal}  &                             \\
Recalibration     &     0.9\%~~~~~~~~~~~2.0\% &      0.9\%~~~~~~~~~~~2.0\%  \\
Efficiency        &      -~~~~~~~~~~~~~~~-    &      5.0\%~~~~~~~~~~~5.0\% \\
Background        &     1.4\%~~~~~~~~~~~2.2\% &      1.9\%~~~~~~~~~~~2.7\% \\
&&\\
                  &~~~~~~~~~~~{\bf Predicted Bremsstrahlung}  &           \\
&&\\
Efficiency        &     5.0\%~~~~~~~~~~~5.0\% &       -~~~~~~~~~~~~~~~-    \\
Formula (2)       &     4.0\%~~~~~~~~~~10.0\% &      4.0\%~~~~~~~~~~10.0\% \\
&&\\
\hline
\end{tabular}
\end{center}

\subsection{Estimation of the accuracy of the bremsstrahlung predictions}

The estimation of the accuracy of the bremsstrahlung predictions resulting 
from formula (2) was done by comparing FSR rates obtained with this formula 
and those delivered by the DYMU3 generator, in the corresponding  $p_T$ ranges,
as the difference between the predictions. In the LE band this uncertainty was
about 4\%, in the HE band about 10\%. They are quoted in Table 1.
 
These estimates agree well with the differences in the predictions for the muon 
inner bremsstrahlung rates obtained with formula (2), and those calculated
with formulae which account for higher order radiative corrections, the
calculations being performed with the non-exponentiated photon spectrum 
\cite{berends2} and with the exponentiated one \cite{yfs,yellow}.
In particular, the latter give 5.9\% and 9.1\% differences with formula (2)
in the LE and HE bands, respectively. Note that 
when doing these calculations, parameter $\beta$ which governs the 
bremsstrahlung photon spectrum \cite{hais} was obtained by integration of 
formula (1.2) in \cite{yfs} applying our $p_T$ cuts, i.e.
within rather narrow angular ranges varying as a function of the photon
energy according to the $p_T$ cuts imposed in the corresponding energy band.  
The $\beta$ values were found to be 0.0146 in the LE band and 0.0088 in the 
HE band, i.e. considerably smaller than 
$\beta = 2 \alpha/\pi ~(\ln s/m_{\mu}^2-1) = 0.0582$,
obtained by integration over all angles. The smallness of $\beta$ reduces
the difference between the bremsstrahlung predictions for the exponentiated 
and non-exponentiated photon spectra.  


\section{ Results}
Photon distributions for $\theta_{\gamma}, ~p_T$ and $p_T^2$ 
are presented both for the data and the background (left panels of 
Figs. 2-5), and for their difference (right panels of the figures).
The latter distributions are accompanied by the calculated bremsstrahlung
spectra according to Eq. (2) shown by triangles.

To quantify  the excess of the data over the background the difference 
between them, which represents the measured muon inner bremsstrahlung,
was integrated in the $p_T$ interval from 0 to 40 MeV/$c$ 
for the photons of the LE band, and from 0 to 80 MeV/$c$ for the photons of
the HE band (these intervals correspond to the filled areas in panels d,f 
of Figs. 2-5), and the values obtained were defined as signals. 
However these $p_T$ cuts were not applied when filling the angular 
distributions displayed in Figs. 2-6 in order to keep these distributions 
unbiased. 

\subsection{Energy band ~~0.2 {\boldmath $ < E_{\gamma} \leq 1$ GeV, 
~~$p_T < $ 40 MeV/$c$}}
Photon distributions, uncorrected and corrected for the photon detection
efficiency, are displayed in Figs. 2 and 3, respectively. 
The results for the signal rate are given in Table 2 together with the 
predictions for the muon bremsstrahlung and their ratios. 

\vskip 0.2cm
{\bf Table 2.} The signal (RD$-$Background), the predicted muon inner 
bremsstrahlung (both in units of $10^{-3} \gamma/\mu$) and their ratios 
in the $p_T < 40$ MeV/$c$ range for the photons from the LE band. 
The first errors are statistical, the second ones are systematic.
\begin{center}
\begin{tabular}{ |c c c| }
\hline
&&\\
Value               &~~~~Data uncorrected for~~~~&~~~~Data corrected for~~~~\\
           &~~~~the detection efficiency~~~~&~~~~the detection efficiency~~~~\\
&&\\
\hline
&&\\
Signal               &~~0.412$\pm 0.048\pm 0.007$~~&~~25.9$\pm 4.0 \pm 1.4$~~\\ 
&&\\
Inner Bremsstrahlung&~~0.388$\pm 0.001\pm 0.025$~~&~~23.30$\pm 0.01\pm 0.93$~~\\
&&\\
Signal/IB            &~~1.06$\pm 0.12\pm 0.07 $~~  &~~1.11$\pm 0.17\pm 0.07$~~\\
&&\\
\hline
\end{tabular}
\end{center}
   
As can be seen from Table 2, the predicted and the measured muon 
bremsstrahlung rates agree well, within the measurement errors.
The small differences in Signal/IB ratios between corrected and uncorrected
data in Table 2 (and Table 3 below) arise from the non-uniformity of the
efficiency reweighting factors.

\subsection{Energy band ~~1 {\boldmath $ < E_{\gamma} \leq 10$ GeV, 
~~$p_T < $ 80 MeV/$c$}}
Photon distributions, uncorrected and corrected for the photon detection
efficiency, are displayed in Figs. 4 and 5, respectively. 
The results for the signal rate are given in Table 3 together with the 
predictions for the muon bremsstrahlung and their ratios.

As can be seen from Table 3, the predicted and the measured muon 
bremsstrahlung rates agree well, within the measurement errors. 
The smaller values of the corrected experimental and predicted bremsstrahlung 
rates in the HE energy band as compared to those in the LE band 
(while the energy range factor following from formula (2),
$\ln(E_{\gamma}^{max}/E_{\gamma}^{min})$, works in favour of the HE band
with an enhancement factor of 1.43) are explained by a higher attenuation 
of the rates induced by the $p_T$ cut in the case of the bremsstrahlung 
photons from the HE band. 

\vskip 0.2cm
{\bf Table 3.} The signal (RD$-$Background), the predicted muon inner 
bremsstrahlung (both in units of $10^{-3} \gamma/\mu$) and their ratios 
in the $p_T < 80$ MeV/$c$ range for the photons from the HE band.
The first errors are statistical, the second ones are systematic. 
\begin{center}
\begin{tabular}{ |c c c| }
\hline
&&\\
Value               &~~~~Data uncorrected for~~~~&~~~~Data corrected for~~~~\\
            &~~~~the detection efficiency~~~~&~~~~the detection efficiency~~~~\\
&&\\
\hline
&&\\
Signal              &0.829$\pm 0.069 \pm 0.025$  &  21.1$ \pm 2.2 \pm 1.3 $ \\ 
&&\\
Inner Bremsstrahlung&0.794$\pm 0.001 \pm 0.089$  &  20.00$ \pm 0.01\pm 2.00$\\
&&\\
Signal/IB           &1.04$ \pm 0.09 \pm 0.12 $   &  1.06$ \pm 0.11 \pm 0.12$ \\
&&\\
\hline
\end{tabular}
\end{center}

\subsection{Observation of the dead cone of the muon bremsstrahlung}
The distributions of the photon production angles with a fine binning (of 
1 mrad bin width) are shown in Figs. 6a,b for the combined sample of the
converted photons from both energy bands. The distribution obtained 
after background subtraction (Fig. 6b) is accompanied by 
the calculated bremsstrahlung points. The displayed measured distributions 
are raw spectra, without any unfolding of the detector angular resolution;
the bremsstrahlung spectra calculated with formula (2) were smeared instead 
by the resolution. We prefer to present the uncorrected measured distributions 
in order to demonstrate the independence of the obtained results on the 
correction procedure. 

As can be seen from the plots, the experimental points follow well the 
predicted bremsstrahlung distribution, showing a turnover at the expected 
bremsstrahlung peak position of 4 mrad. This is therefore an observation 
of the muon inner bremsstrahlung dead cone, for the first time 
in high energy physics experiments. The observation
enriches the agreement between the experimental findings of the muon inner 
bremsstrahlung characteristics reported in this work and the QED predictions 
for the process. 

However a deeper insight into the bremsstrahlung pattern can be obtained 
when considering, instead of the distribution $dN_\gamma/d\theta_\gamma$, 
the distribution $dN_\gamma/d\Omega$, where $d\Omega$ is a solid angle
element. Such a distribution is free of kinematic suppression at the polar 
angles $\theta_\gamma$ approaching zero, and the remaining suppression of
the photon production rate at very small angles is a purely dynamic effect,
similar to that mentioned in Sect. 2 for the hadrons inside a jet, namely
a destructive interference between the radiation sources, but this time 
less straightforward, just between the muon ``before" and ``after" the photon 
emission \footnote{In classical language, the radiation intensity into the 
solid angle $d\Omega$ vanishes when three vectors: the muon velocity, its 
acceleration, and the radiation unit vector happen, in particular, to be 
parallel, see for example \cite{ll,jack}.

}. 
The solid angle element $d\Omega$ is proportional to 
$d\cos \theta_\gamma$, which at small angles 
is, in turn, proportional to $d\theta_\gamma ^2$. The position of the
$dN_\gamma/d\theta_\gamma ^2$ distribution turnover is predicted to be at 
$\theta_\gamma ^2 = 1/\Gamma ^2$ ($\Gamma = 430$, see 
Sect. 2), i.e. at $\theta_\gamma ^2 = 5.4 \times 10^{-6}$ rad$^2$. 

To observe this turnover, an improved angular resolution was required,
achieved with the additional cuts (see Sect. 4.2) to
be at the level of 1.4 mrad, as noticed in Sect. 3.2. The distribution 
$dN_\gamma/d\theta_\gamma ^2$ obtained with this resolution is shown in 
Fig. 6c, together with the bremsstrahlung predictions for this variable. 
Though the statistics are poor, the dip at $\theta_\gamma ^2 < 5\times 10^{-6}$ 
rad$^2$ is visible in this distribution, revealing the dynamical dead cone 
of the muon inner bremsstrahlung.    
 
In order to estimate the statistical significance of this observation the 
following procedure was undertaken. The initial part (about 20 bins) of the 
bremsstrahlung $\theta_\gamma ^2$ distribution shown in Fig. 6c, with first two
bins omitted, was fitted by a smooth curve (by a polynomial of 4th or 5th 
order). Then the fitting curve was extrapolated to zero, as shown in the figure 
giving the value of $(5.64 \pm 0.27)~\gamma/5\times 10^{-6}$ rad$^2$ 
at the centre of the first bin of the distribution (the error reflects the 
variation in the fitting form and in the number of bins used in the fit). 
This value was assumed to represent the expected bremsstrahlung rate in the 
first bin of the distribution in a hypothetical situation when the 
bremsstrahlung dynamical dead cone is absent.
The number of the real data photons in the first bin was 2 with the estimated
background to be $0.66 \pm 0.46$, thus giving the signal value in this bin 
$(1.34 \pm 1.49)~\gamma /5\times 10^{-6}$ rad$^2$. 
Assuming Poisson distribution for the signal photons these numbers correspond
to the probability of the absence of the bremsstrahlung dead cone of less 
than 4\%.
  
\section{Comparison with the hadronic soft photon analysis}
The main difference between the results of this analysis and the hadronic 
ones \cite{aspdel} is the absence of any essential excess of the soft photon 
production over the predicted inner bremsstrahlung rate reported in this 
study, contrary to the case for \cite{aspdel} where the observed soft 
photon rate was found to exceed the bremsstrahlung predictions by a factor
of about 4. The 95\% CL upper limits on the excess factors which can be 
extracted from the results of this work are 1.29 in the LE band, and 1.28 
in the HE band.  
     
Another distinction between the two analyses is an essential difference
in the background levels and in the possible systematic effects. However,
the code transporting photons through the DELPHI detector and simulating
their conversions in the detector material (DELSIM), 
the photon reconstruction algorithm and the determination of its 
efficiency, together with the recalibration procedure, were common
to the two analyses. Thus the results of this work can be considered also 
as a cross-check of these procedures in the hadronic events study. On the 
other hand, the amount of dimuon events collected during the LEP1 period is 
considerably smaller than the number of collected hadronic events, due to 
a smaller $Z^0$ dimuon branching ratio (by a factor of 20). As a result, 
in the current analysis the statistical errors are either essentially higher 
than the systematic ones (in the LE band), or comparable to them  (in the HE 
band), while in \cite{aspdel} the total uncertainties of the measured photon 
rates are dominated by systematic errors; nevertheless it should be emphasized 
that the results of both analyses show clear signals of direct photons 
(even though the strength of the signal in \cite{aspdel} is 
not explained theoretically). 

\section{Summary}   
The results of the analysis of final state radiation in $\mu^+ \mu^-$ 
decays of $Z^0$  events at LEP1 are reported in this work. 
The radiation was studied in the region of small transverse momenta with 
respect to the parent muon, $p_T < 40$ MeV/$c$ in the photon energy range 
$0.2 < E_\gamma \leq 1$ GeV (LE band), and  $p_T < 80$ MeV/$c$ in the photon 
energy range $1 < E_\gamma \leq 10$ GeV (HE band). 
The obtained photon rates uncorrected (corrected) for the photon detection 
efficiency were found to be, in units of  $10^{-3} \gamma/\mu$, with the 
first error to be statistical and the second one systematic: a) in the 
LE band: 0.412$ \pm 0.048 \pm 0.007 $ (25.9$\pm 4.0  \pm 1.4$), 
while QED predictions for the muon inner bremsstrahlung were calculated to be 
0.388$ \pm 0.001 \pm 0.025$ (23.30$\pm 0.01 \pm 0.93$); 
b) in the HE band: 0.829$\pm 0.069 \pm 0.025$ (21.1$ \pm 2.2 \pm 1.3 $), 
while the muon inner bremsstrahlung was calculated to be 
0.794$\pm 0.001 \pm 0.089$ (20.00$ \pm 0.01\pm 2.00$).
The obtained ratios of the observed signal to the predicted level of the
muon inner bremsstrahlung are then $1.06 \pm 0.12 \pm 0.07$ in the LE band 
and $1.04 \pm 0.09 \pm 0.12$ in the HE band (uncorrected rates are used for 
these ratios, as they possess a better statistical accuracy). Thus, the 
analysis shows a good agreement between the observed photon production rates 
and the QED predictions for the muon inner bremsstrahlung, both in differential
(see Figs. 2-5) and integral (see Tables 2,3) forms. This is in
contrast with the anomalous soft photon production in hadronic decays of $Z^0$ 
reported earlier in \cite{aspdel}.

The bremsstrahlung dead cone is observed for the first time in
the direct photon production in $Z^0$ decays in particular, and in
the muon inner bremsstrahlung in the high energy physics experiments in 
general, also being in good agreement with the predicted bremsstrahlung 
behaviour.

\section*{Acknowledgements}
We thank Profs. K. Boreskov, J.E. Campagne, F. Dzheparov and A. Kaidalov 
for useful discussions.

We are greatly indebted to our technical
collaborators, to the members of the CERN-SL Division for the excellent
performance of the LEP collider, and to the funding agencies for their
support in building and operating the DELPHI detector.\\
We acknowledge in particular the support of \\
Austrian Federal Ministry of Education, Science and Culture,
GZ 616.364/2-III/2a/98, \\
FNRS--FWO, Flanders Institute to encourage scientific and technological
research in the industry (IWT) and Belgian Federal Office for Scientific,
Technical and Cultural affairs (OSTC), Belgium, \\
FINEP, CNPq, CAPES, FUJB and FAPERJ, Brazil, \\
Ministry of Education of the Czech Republic, project LC527, \\
Academy of Sciences of the Czech Republic, project AV0Z10100502, \\
Commission of the European Communities (DG XII), \\
Direction des Sciences de la Mati$\grave{\mbox{\rm e}}$re, CEA, France, \\
Bundesministerium f$\ddot{\mbox{\rm u}}$r Bildung, Wissenschaft, Forschung
und Technologie, Germany,\\
General Secretariat for Research and Technology, Greece, \\
National Science Foundation (NWO) and Foundation for Research on Matter (FOM),
The Netherlands, \\
Norwegian Research Council,  \\
State Committee for Scientific Research, Poland, SPUB-M/CERN/PO3/DZ296/2000,
SPUB-M/CERN/PO3/DZ297/2000, 2P03B 104 19 and 2P03B 69 23(2002-2004),\\
FCT - Funda\c{c}\~ao para a Ci\^encia e Tecnologia, Portugal, \\
Vedecka grantova agentura MS SR, Slovakia, Nr. 95/5195/134, \\
Ministry of Science and Technology of the Republic of Slovenia, \\
CICYT, Spain, AEN99-0950 and AEN99-0761,  \\
The Swedish Research Council,      \\
Particle Physics and Astronomy Research Council, UK, \\
Department of Energy, USA, DE-FG02-01ER41155, \\
EEC RTN contract HPRN-CT-00292-2002. \\


\newpage
\begin{figure}[1]
\begin{center}
\epsfig{file=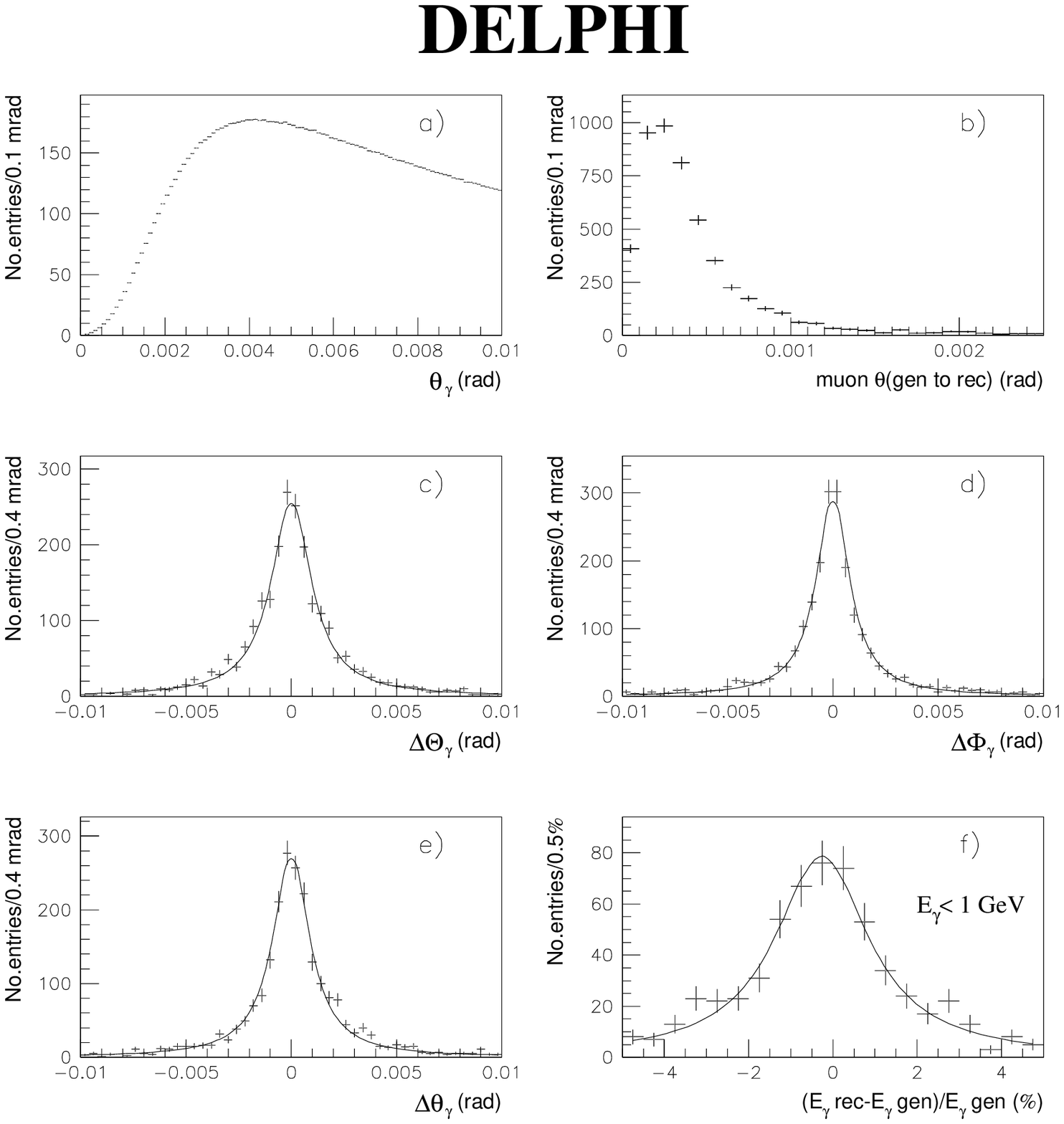,bbllx=50pt,bblly=180pt,bburx=550pt,bbury=570pt,%
width=17cm,angle=0}
\end{center}
\caption{ a) Distribution of the final state radiation production angle in 
$Z^0 \rightarrow \mu^+ \mu^-$ events generated with formula (2); b) opening 
angle between the generated and reconstructed directions of a muon track; 
c) difference between generated and reconstructed photon polar angles 
$\Theta_\gamma$; d) the same for the azimuthal angles $\Phi_\gamma$; 
e) difference between generated and reconstructed $\mu \gamma$ opening angles 
$\theta_{\gamma}$, which illustrates the overall angular resolution of 
this analysis;
f) difference between the generated and the reconstructed photon energies 
in the photon energy range of $0.2 < E_\gamma \leq 1$ GeV. The curves in
Fig. 1c-1f are the fits by Breit-Wigner forms (see text).} 
\end{figure}
\newpage
\begin{figure}[2]
\begin{center}
\epsfig{file=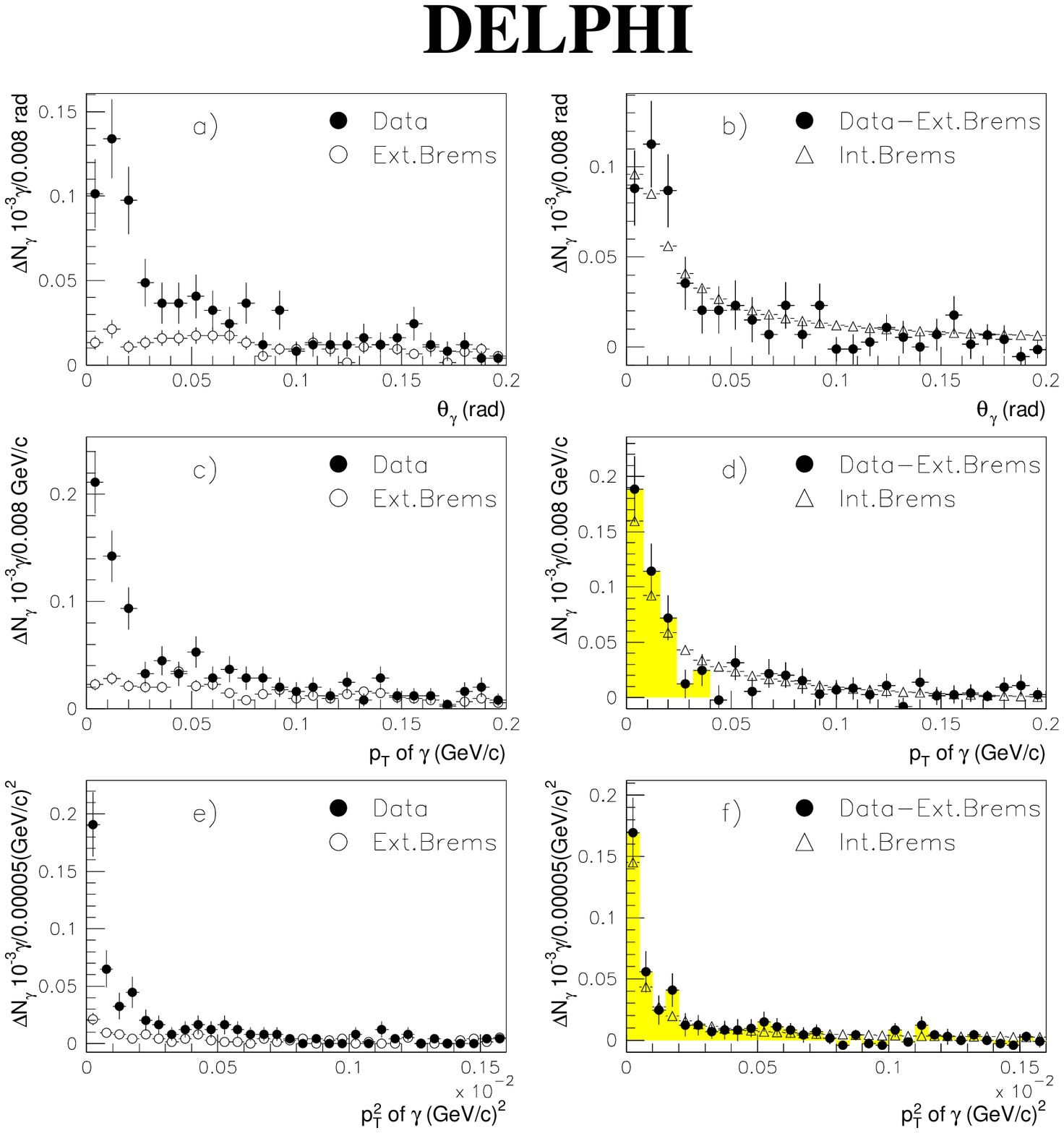,bbllx=50pt,bblly=180pt,bburx=550pt,bbury=570pt,%
width=17cm,angle=0}
\end{center}
\caption{Photon distributions in the photon energy band $0.2 < E_\gamma \leq 1$
GeV uncorrected for the photon detection efficiency. 
The photon rates are given as the number of photons per 1000 muons
per bin width of the distribution.
Left panels: the data and background distributions for a)
$\theta_{\gamma}$, the photon production angle;
c) photon $p_T$; e) photon $p_T^2$. Right panels, b), d), f): the difference
between the data and the background for the same variables, respectively.
``Ext.Brems" corresponds to the background,
``Int.Brems" corresponds to the muon inner bremsstrahlung predictions.
The filled areas in panels d) and f) correspond to the signal integral
(see text). The errors shown are statistical.} 
\end{figure}
\begin{figure}[3]
\begin{center}
\epsfig{file=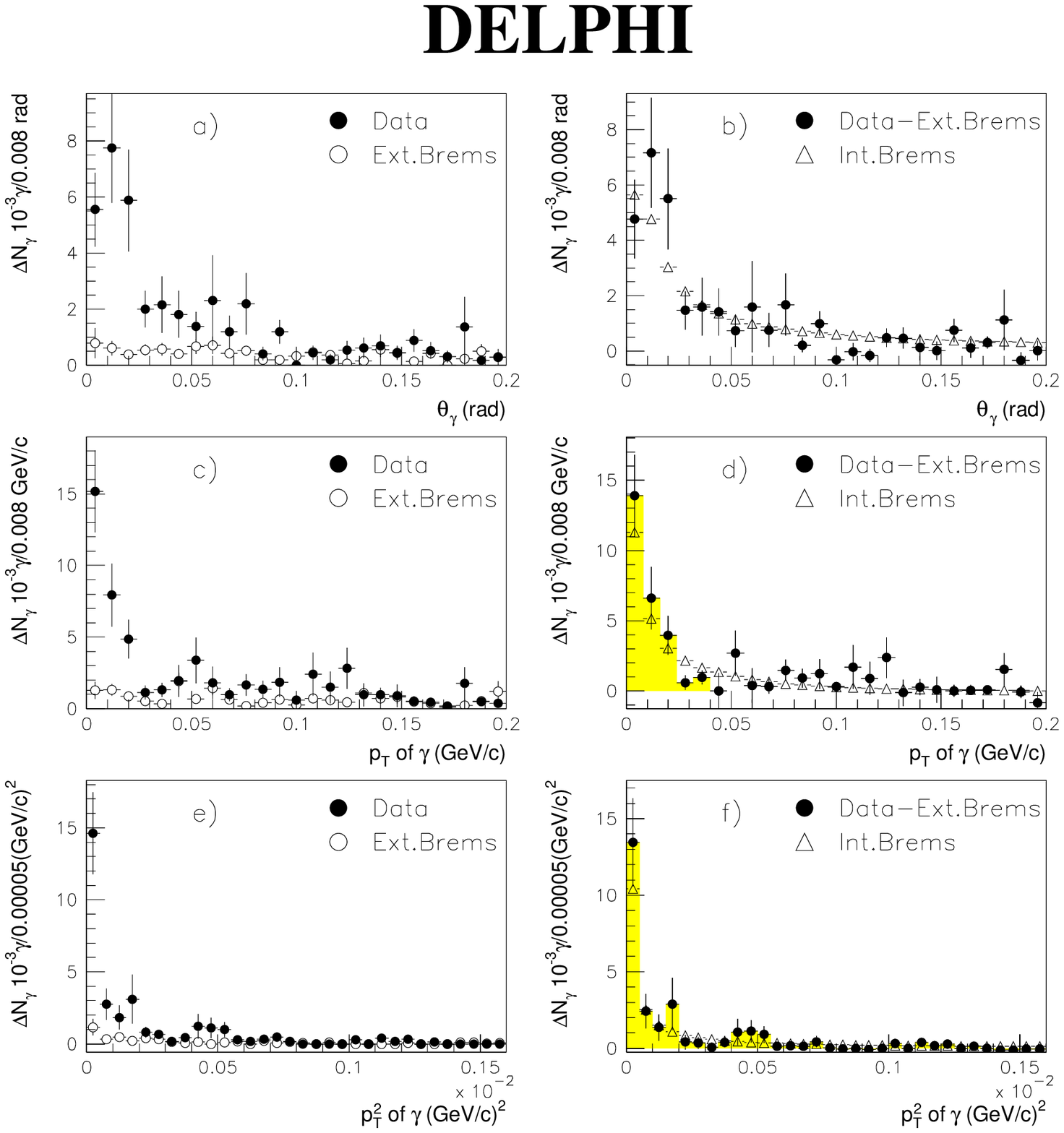,bbllx=50pt,bblly=180pt,bburx=550pt,bbury=550pt,%
width=17cm,angle=0}
\end{center}
\caption{The same as in Fig. 2, corrected for the efficiency of photon
detection.}
\end{figure}

\begin{figure}[4]
\begin{center}
\epsfig{file=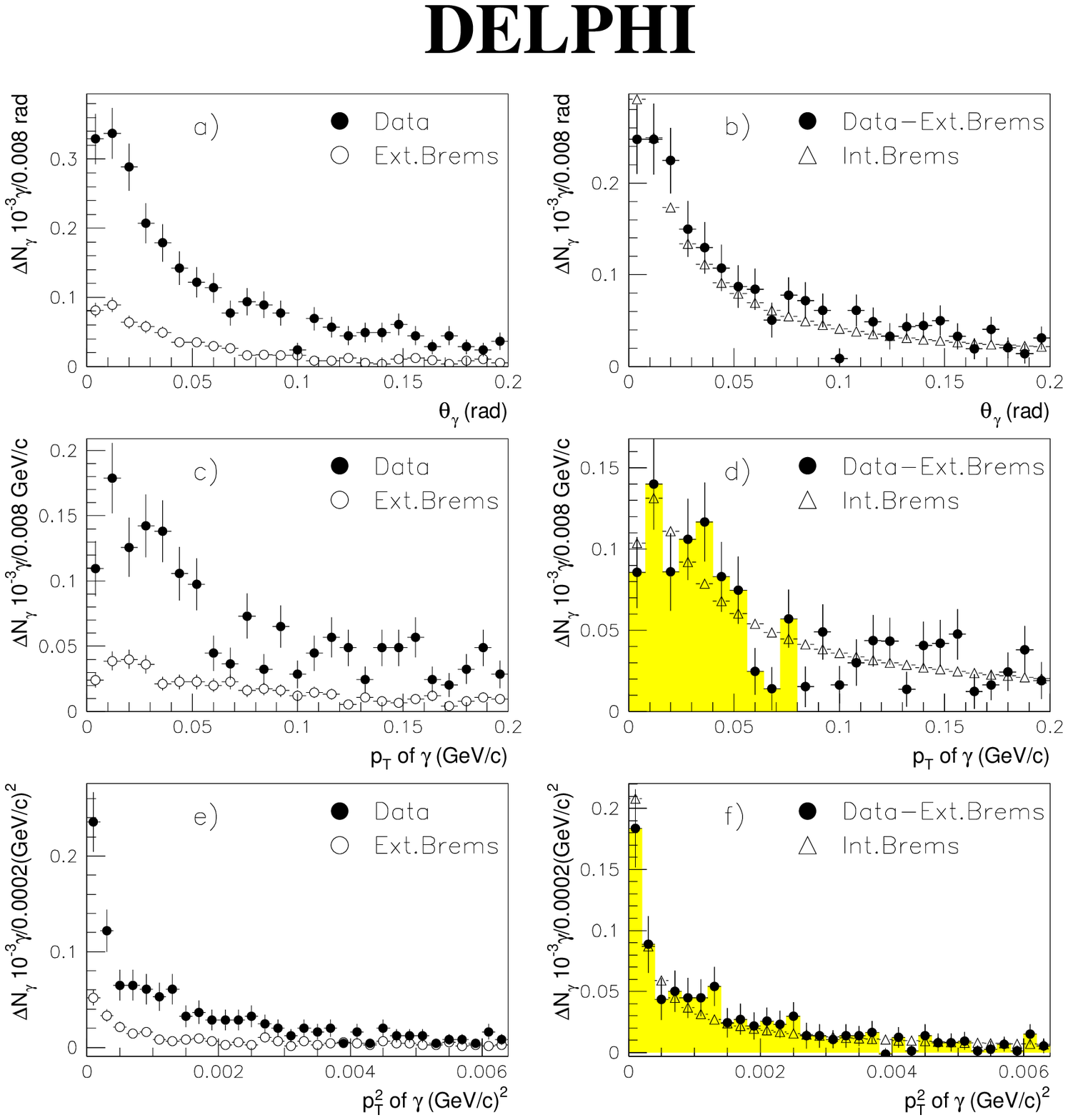,bbllx=50pt,bblly=180pt,bburx=550pt,bbury=570pt,%
width=17cm,angle=0}
\end{center}
\caption{Photon distributions in the photon energy band $1 < E_\gamma \leq 10$ 
GeV uncorrected for the photon detection efficiency. The photon rates are
given in number of photons per 1000 muons per bin width of the distribution. 
Left panels: the data and background distributions for a)
$\theta_{\gamma}$, the photon production angle;
c) photon $p_T$; e) photon $p_T^2$. Right panels, b), d), f): the difference
between the data and the background for the same variables, respectively.
The filled areas in panels d) and f) correspond to the signal integral
(see text). The errors shown are statistical.} 
\end{figure}

\begin{figure}[5]
\begin{center}
\epsfig{file=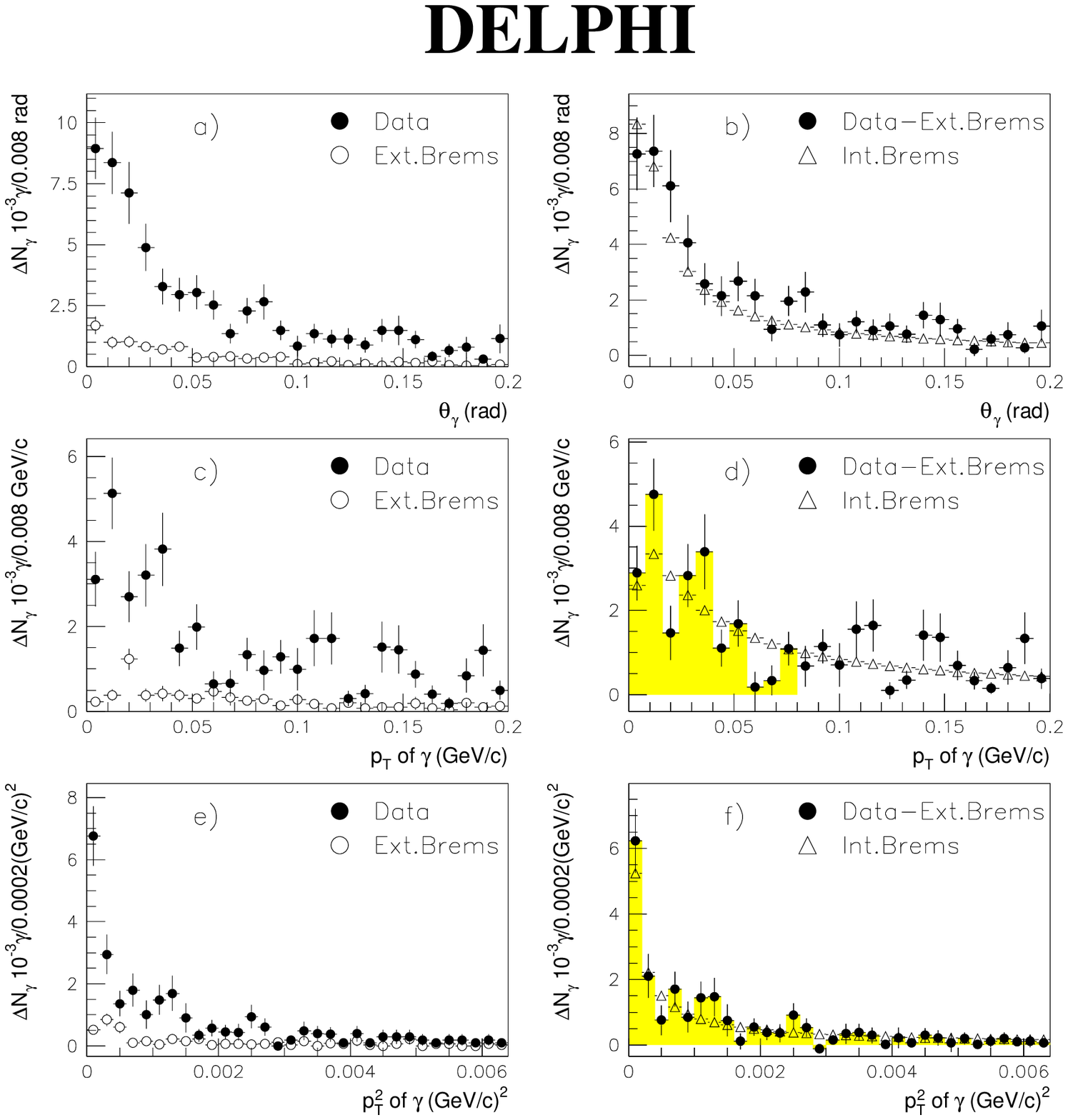,bbllx=50pt,bblly=180pt,bburx=550pt,bbury=550pt,%
width=17cm,angle=0}
\end{center}
\caption{The same as in Fig. 4, corrected for the efficiency of photon
detection.}
\end{figure}

\begin{figure}[6]
\begin{center}
\epsfig{file=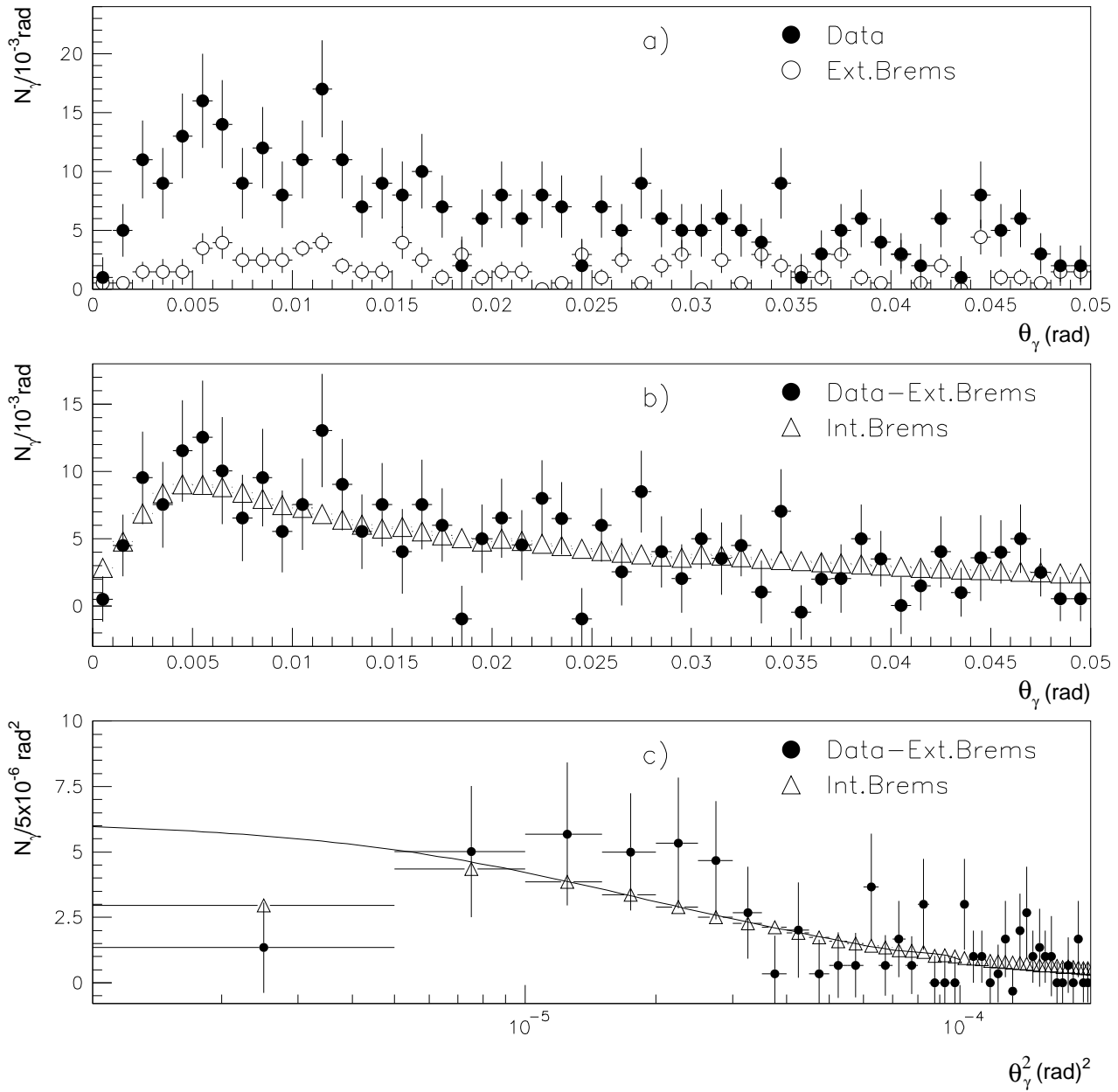,bbllx=50pt,bblly=180pt,bburx=550pt,bbury=550pt,%
width=17cm,angle=0}
\end{center}
\caption{Dead cone of the muon inner bremsstrahlung: a,b) as seen in 
the photon production angle distributions: a) the data and the 
background distributions; b) the difference between the data 
and the background; and c): the distribution of the photon production angle 
squared, obtained under tighter cuts that improve the angular resolution;
the curve shows the fit of the bremsstrahlung distribution within 
$10^{-5} \leq \theta_\gamma ^2 < 10^{-4}$ rad$^2$ by a 5th order polynomial 
extrapolated to the 1st bin of the distribution (see text). 
The errors shown are statistical.}
\end{figure}

\end{document}